\documentclass[12pt]{revtex4}
\usepackage{graphics,epsfig,graphicx}
\usepackage{color}
\usepackage{makeidx}
\usepackage{latexsym}
\usepackage[latin1]{inputenc}

\begin{document}


\title{Binding energy of the Wannier exciton on an organic quantum
wire}

\author{ François Dubin$^{1}$ and Michel Schott \\
{\em Institut des Nanosciences de Paris, UMR 7588 of CNRS,} \\
{\em Universit\'{e} Pierre et Marie Curie et Universit\'{e} Denis Diderot,\\
Campus Boucicaut, 140 rue de Lourmel}  {\em 75015 Paris}}

\begin{abstract}
The exciton on a single polydiacetylene chain is considered in a
Wannier approach, taking into account the surrounding polarizable
medium. The electron-hole Coulomb interaction potential is
explictly obtained for a quantum wire of circular cross-section,
and numerically calculated for a rectangular one. The predicted
binding energy and Bohr radius, starting from the chain
conformation, are in good agreement with experimental values. This
shows the importance of considering the monomer matrix embedding
the chain, and the major role of confinement, together with
electron correlations, for describing the exciton. The method is
general and applicable to any quantum wire.
\end{abstract}

\maketitle


A large amount of experimental and theoretical works is currently
done on excitons in semiconductor quantum wires (QWRs). Excitonic
properties of strictly one-dimensional (1D) systems are known to
be singular, with an infinite ground state exciton binding energy
and a zero Bohr radius. However, any physical wire has finite
lateral dimensions; once this is taken into account, the strictly
1D pathological behaviour disappears \cite{Monique1}.

We have developed organic polymeric quantum wires which show the
theoretically expected - but up to now rarely observed- behaviour
for quasi-one dimensional excitons, i.e., a radiative lifetime
increasing as $\sqrt{T}$ \cite{Lecuiller2002}, and a 1/$\sqrt{E}$
energy singularity at the band edge \cite{Dubin2002}. These wires
consist of a polydiacetylene (PDA) chain, made of a linear chain
of Carbon atoms linked by alternating single, double, and triple
bonds. PDA's are typical conjugated polymers, and can be very
diluted in their single crystal monomer matrix such that
interchain interactions are negligible, the inhomogeneous
broadening being then minimal, with each chain possibly considered
as an isolated QWR. Its lateral dimensions are of the order of the
atomic radius of Carbon, i.e. a few $\AA$; This is clearly close
to the lowest achievable limit.

The experimentally measured binding energy of PDA ground state
exciton is quite large, E$_b$= 0.6 eV \cite{Horvath}, and the
inferred exciton radius fairly small, a$_{1D} \approx$ 20 $\AA$
\cite{Moller}. Nevertheless, the exciton is not localized by
disorder since the exciton resonance emission energy is the same
at all positions along a 20 $\mu$m long chain  within the
experimental resolution of 50 $\mu$eV \cite{dubin_nature}.

There are indications that such chains are strongly correlated systems
\cite{Jeckelmann}, so the large E$_b$ may be associated to lateral
confinement, or to correlations, or to both. The purpose of this work is to
explore the role of lateral confinement alone, by considering the polymeric
wire as an ordinary semiconductor QWR (as in \cite{Monique1}).

A straightforward calculation of a QWR having the PDA chain
conformation in vacuum leads to a too large E$_b$ and too small
a$_{1D}$. Quantum chemical calculations on PDA's or other
conjugated polymers meet the same problem, with predicted E$_b$ of
the order of 2 eV \cite{VderHorst}. The reason is obvious: PDA
chains in our experiments - and in all similar studies on
conjugated polymers- are embedded in a dielectric, polarizable,
medium - the monomer single crystal - which influence on the
electronic energy levels must be taken into account, as was
already done for molecular crystals. This is done in the present
paper, the effect being computed within the standard image method
\cite{2dci,Durand}.

Based on the treatment given in \cite{Monique1}, the appropriate
electron-hole Coulomb interaction potential for quasi-1D ground
state excitons is first recalled in section I, and the associated
E$_b$ and a$_{1D}$ are deduced. In section II, the influence of
the surrounding medium on the ground state excitonic parameters is
introduced. By the mean of appropriate dielectric constants,
$\epsilon_w$ and $\epsilon_m$, for the wire and medium
respectively, the electron-hole interaction potential is
rewritten. Its exact evaluation being somewhat complex, we propose
an alternative procedure which is approximate but quite precise.
It is based on the image charges of the confined carriers. This
treatment yields a much simpler expression for the electron-hole
interaction potential. It is used in section III to calculate
E$_b$ and a$_{1D}$ for PDA´s. The accuracy of our model is
demonstrated in the last section by comparing the obtained values
of E$_b$ and a$_{1D}$ to the measured ones. It is of importance to
stress that the proposed method is actually quite general and not
restricted to organic quantum wires.

\section{Quasi-1D electron-hole Coulomb potential}\label{I}

We consider a QWR with a rectangular section along (Oxy) and a
free axis along (Oz). Let ($x_{e},y_{e},z_{e}$) and
($x_{h},y_{h},z_{h}$) be the coordinates of the electron and hole
confined in the wire. As shown in \cite{Monique1}, the exact
interaction potential between the electron and the hole for the
lowest energy singlet exciton, so called ground state exciton in a
semiconductor terminology (X$_0$), can be written in terms of
$z$=($z_{e}-z_{h}$) only. It reads
\begin{equation}\label{eq1}
V(z_e-z_h)= \left< f_{h,1}\left| \bigotimes \left< f_{e,1}\left|
\frac{-e^2}{\sqrt{(z_e-z_h)^2+(x_{e}-x_{h})^2+(y_{e}-y_{h})^2}}\right|
f_{e,1} \right>\bigotimes \right| f_{h,1}\right>
\end{equation}
where $f_{e,1}$ ($f_{h,1}$) is the eigen-function for the first energy level
of the 2D confined motion of the free electron (hole) in the (Oxy) plane (see
for instance Appendix A of \cite{Monique1}).

In order to find an analytical expression for the X$_0$ relative
motion wave function, Combescot et al. have shown that
$V(z_e-z_h)$ can be approximated by
$V_{eff}(z_e-z_h)=-e^{2}/(|z_e-z_h| +b^*)$ where $b^*$ represents
the effective width of the wire. It is such that
$V_{eff}(0)=-e^{2}/b^*$. By setting $b^*=\lambda
a_{x}\beta_{\lambda}^{*}/2$, $a_{x}$ being the 3D exciton Bohr
radius and $\lambda$ the Landau re-scaling energy parameter, the
X$_0$ relative motion wave function reads
\begin{equation} \phi_{\lambda,0}(z >
0)=Ae^{-(z+\beta^{*}_{\lambda})/2}U(-\lambda,0,z+\beta_{\lambda}^{*})
\end{equation}
\begin{equation}
\phi_{\lambda,0}(z < 0)
=Ae^{-(-z+\beta^{*}_{\lambda})/2}U(-\lambda,0,-z+\beta_{\lambda}^{*})
\end{equation}
where $U$ denotes the Kummer series \cite{Bateman}. The
eigenvalues $\lambda$ are given by the continuities of the wave
function and its derivative for $z=z_e-z_h=0$.

The effective potential built with this $b^*$ has to be good over the ground
state exciton extension. Therefore, this $b^*$ parameter needs to be chosen
very carefully since E$_b$ strongly depends on it. At first order, $b^*$ is
selfconsistently obtained from $\lambda$ through
\begin{equation}\label{eq4}
D(\lambda)=\left<\phi_{\lambda,0}\left|V_{eff}(z)+\frac{e^2}{|z|+\lambda
a_{x}\beta_{\lambda}^{*}/2}\right|\phi_{\lambda,0} \right>=0
\end{equation}

Consequently, we have to look for the appropriate energy parameter
$0<\lambda_{0}<1$ such that $D(\lambda_0)$=0, and then, deduce the $X_0$
binding energy E$_b$=(-$R_{x}/\lambda_{0}^2$), $R_{x}$ being the 3D Rydberg.
The effective $X_0$ Bohr radius is obtained through
$a_{1D}=\sqrt{<\phi_{\lambda_0,0}|z^2|\phi_{\lambda_0,0}>}$.

\section{Polarizability of the quantum wire outside medium}\label{I}

The exact electron-hole interaction potential presented above
(equation (\ref{eq1})) is obtained assuming that the QWR
dielectric constant, $\epsilon_w$, is equal to the one of the
outside medium, $\epsilon_m$. In this case, by construction, the
potentials created by the confined carriers are continuous at the
boundary between the QWR and its environment. On the contrary, if
($\epsilon_w \neq \epsilon_m$) equation (1) is no more valid. The
potentials created by the confined electron and hole both need to
be rewritten in order to ensure their continuity at the boundary
between the QWR and the outside. The latter modifies the
electron-hole Coulomb interaction potential. To find its exact
expression for the $X_0$, $V^{(\epsilon_w \neq \epsilon_m)}$, a
circular wire has to be studied \cite{Durand}. Actually, circular
and rectangular wires have been shown to give identical $X_0$
parameters if their cross sections are the same \cite{Monique1}.

\subsection{Exact potential for a circular wire}

In order to derive the equivalent of equation (1) for $(\epsilon_w \neq
\epsilon_m)$, we first establish the expression of the potential created by
an electron confined within the wire, $V^{(\epsilon_w \neq \epsilon_m)}_{e}$.
Let $R$ be the wire radius, and $(\rho_{e},\theta_{e},z_{e})$ the electron
position in polar coordinates. $V^{(\epsilon_w \neq \epsilon_m)}_{e}$ reads
in terms of the Bessel functions $I_0$ and $K_0$ \cite{Durand} as
\begin{eqnarray}
V^{(\epsilon_w \neq
\epsilon_m)}_{e,in}(\rho,\theta,z)=\frac{-2e}{\pi}\int_{0}^{\infty}[K_{0}(m\|\rho-\rho_{e}\|)+A(m,\rho_{e})I_{0}(m\|\rho-\rho_{e}\|)] \nonumber\\
cos(m(z-z_{e}))dm\nonumber \\
V^{(\epsilon_w \neq
\epsilon_m)}_{e,out}(\rho,\theta,z)=\frac{-2e}{\pi}\int_{0}^{\infty}B(m,\rho_{e})K_{0}(m\|\rho-\rho_{e}\|)\nonumber\\
cos(m(z-z_{e}))dm\label{Vint}
\end{eqnarray}
$V^{(\epsilon_w \neq \epsilon_m)}_{e,in}$ and $V^{(\epsilon_w \neq
\epsilon_m)}_{e,out}$ being the expression of $V^{(\epsilon_w \neq
\epsilon_m)}_{e}$ inside and outside the wire respectively. The functions $A$
and $B$ appearing in (\ref{Vint}) are deduced from the two boundary
conditions
\begin{equation} \label{c_limites_V}
(V^{(\epsilon_w \neq \epsilon_m)}_{e,in}- V^{(\epsilon_w \neq
\epsilon_m)}_{e,out})_{\rho=R}=0, \hspace{.3cm}
\left(\epsilon_{w}\frac{\partial V^{(\epsilon_w \neq
\epsilon_m)}_{e,in}}{\partial\rho}-\epsilon_{m}\frac{\partial V^{(\epsilon_w
\neq \epsilon_m)}_{e,out}}{\partial\rho}\right)_{\rho=R}=0
\end{equation}
Consequently, from equations (\ref{eq1}-\ref{Vint}-\ref{c_limites_V}), we
find that equation (1) transforms into
\begin{equation}
V^{(\epsilon_c \neq \epsilon_m)}(z_e-z_h)=e<f_{h,1} \mid \otimes < f_{e,1}
\mid V^{(\epsilon_w \neq \epsilon_m)}_{e,in} (\rho_h,\theta_h,z_h)\mid
f_{e,1}>\otimes\mid f_{h,1}>\label{V_polaire}
\end{equation}

The evaluation of (\ref{V_polaire}) yields a fifth integration of
an oscillating product of Bessel functions over an infinite
domain. This integral obviously converges, but is technically
difficult to compute. To calculate the electron-hole interaction
potential we present an alternative procedure, approximate but
precise enough, based on the electron-hole charge images
\cite{2dci,Durand}. For that purpose, let us first recall the
basic results of the charge image representation.

\subsection{Image charges for a rectangular wire}

\begin{figure}
\includegraphics[width=10cm]{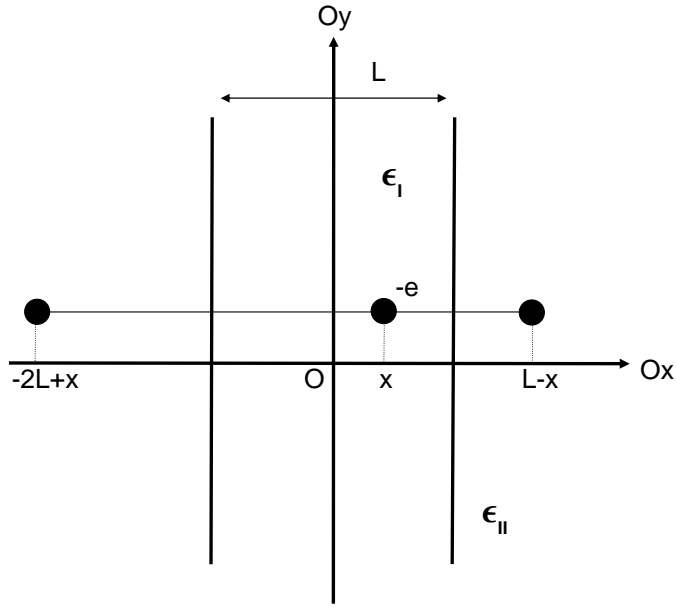}
\caption{Image distribution for a charge $q$=-$e$ confined in a
dielectric medium (I) separated from its environment (II) by two
infinite planes, as for a charge in a finite quantum well. The
image charges in (II) need to be considered to evaluate the
potential created by the confined carrier $q$.}
\end{figure}

We consider two media, (I) and (II), with dielectric constants
$\epsilon_I$ and $\epsilon_{II}\neq\epsilon_I$. Let (I) be
embedded in (II) and separated by planar interfaces. The potential
created by a charge $q$ confined in (I) is strongly affected by
the dielectric mismatch ($\epsilon_I\neq\epsilon_{II}$). This
potential can be expressed in terms of the electrostatic images of
$q$ in (II). With
$S=(\epsilon_I-\epsilon_{II})/(\epsilon_I+\epsilon_{II})$, a
charge $q$ located at $(x_q,y_q,z_q)$ in (I) has a double set of
($S^{|n|}q)$ electrostatic images in (II), located at ($(-1)^{n}(
x_q-nL),y_q,z_q)$ (see figure 1). The charge $q$ plus its images
lead a potential in the medium (I) which reads
\begin{eqnarray}
V_q^{(I)}(x,y,z)=\frac{q}{\sqrt{(x-x_{q})^2+(y-y_{q})^2+(z-z_q)^2}}\nonumber\\
+\sum_{n\neq
0}\frac{S^{|n|}q}{\sqrt{(x-(-1)^n(x_q-nL))^2+(y-y_{q})^2+(z-z_q)^2}}
\end{eqnarray}

This electrostatic images representation has been succesfully used
to calculate the binding energy of an exciton confined in a
quantum well close to a metallic mirror as well as for a quasi-2D
electron gas confined in a finite quantum well \cite{2dci}. For
carriers in quantum wells the procedure is rather simple since the
interfaces between the well and the outside medium are made of two
2D planes (see fig. 1). On the contrary, the case of charges in
quantum wires is far more subtle. Besides the fact that the QWR
cross section is very small, the inclusion of the dielectric
mismatch between the wire and its outside medium in terms of
electrostatic images is questionable. We nevertheless apply it in
the following, and show in section III that this method leads to
relevant results, as many similar ones in which bulk properties
are extended to rapidly varying situations.

\begin{figure}
\includegraphics[width=10cm]{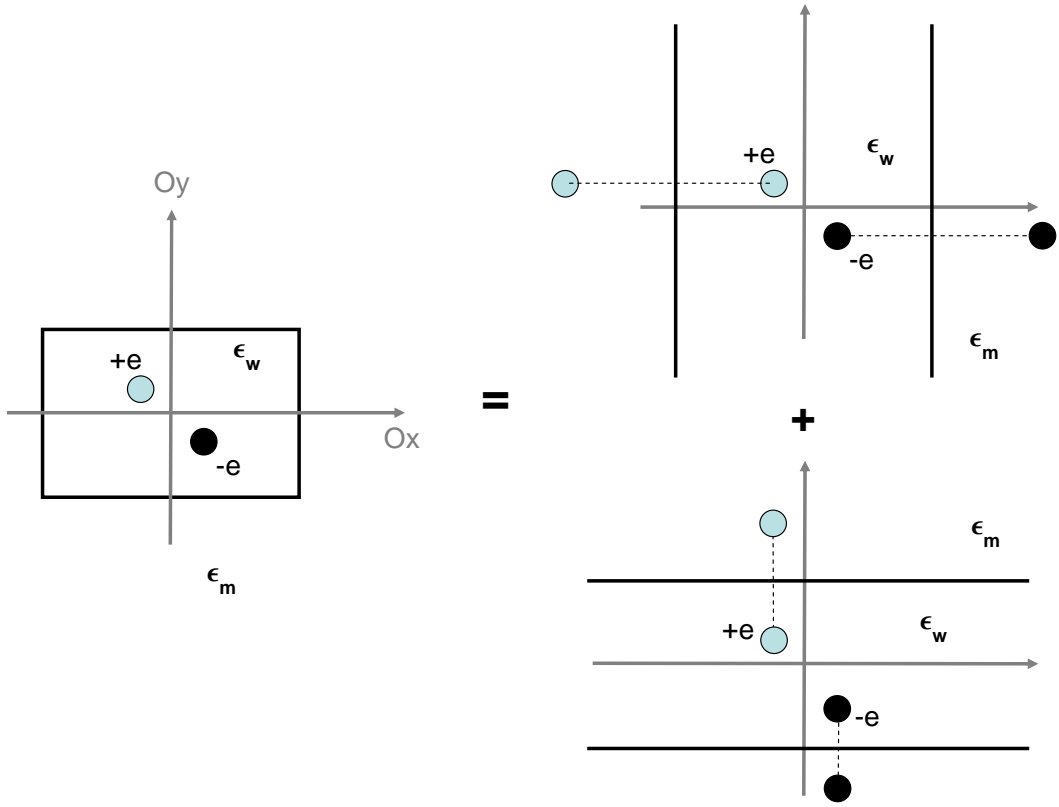}
\caption{Image distribution of an electron-hole pair confined in a
rectangular quantum wire. Their interaction potential is evaluated
with the sum of the vertical and horizontal images of the electron
and hole (see equation (12)).}
\end{figure}

Let us consider a QWR of dielectric constant $\epsilon_w$,
embedded in a medium of dielectric constant $\epsilon_m$. The wire
section is rectangular with extension $L_{x}$=a along (Ox) and
$L_{y}$=b along (Oy). Let
$T=(\epsilon_w-\epsilon_m)/(\epsilon_w+\epsilon_m)$ and
$(x_e,y_e,z_e)$ ( with $|x_e|<$a/2 and $|y_e|<$b/2) be the
electron coordinates in the wire. The carrier has then two
infinite sets of (-$T^{|n|}e)$ electrostatic images, one along
(Ox), placed at ($(-1)^{n}(x_e-na),y_e,z_e)$, and another set of
images along (Oy), having the same (-$T^{|n|}e)$ charge and placed
at $(x_e,(-1)^{n}(y_e-nb),z_e)$, $n$ being a positive or negative
integer.  From the distribution of the electron images, the
potential it creates, $V_{e}^{(im)}$, inside and outside the wire,
$V_{e,in}^{(im)}$ and $V_{e,out}^{(im)}$ respectively, can be
expressed. $V_{e,out}^{(im)}$ has a form depending on the region
where it is calculated. For instance, on the right side of the
wire, i.e., at $(x>$a/2$,y,z)$, only images on the left side of
the wire, i.e., having an abscissa $(x<$-a/2), have to be
included. $V_{e,out}^{(im)}$ there reads
\begin{eqnarray}
V^{(im)}_{e,out}(x,y,z)=\frac{-Te}{\sqrt{(x-x_{e})^2+(y-y_{e})^2+(z-z_e)^2}}\nonumber\\
-\sum_{n>0}\frac{-T^{|n|}e}{\sqrt{(x-(
x_{e}-na))^2+(y-y_{e})^2+(z-z_e)^2}}, \hspace{0.1cm}
for\hspace{0.1cm} x>a/2
\end{eqnarray}
The expression of $V_{e,out}^{(im)}$ at a random $(x,y,z)$ position is
deduced generalizing the previous example \cite{Durand}. It is not
extensively written here since this expression is not essential for the
problem studied. On the contrary, $V_{e,in}^{(im)}$ is always given by
\begin{eqnarray}
V^{(im)}_{e,in}(x,y,z)=\frac{-e}{\sqrt{(x-x_{e})^2+(y-y_{e})^2+(z-z_e)^2}}+V^{(im)}_{e_{in}}(x,y,z)\label{eq5}\\
V^{(im)}_{ e_{in}}(x,y,z)=\sum_{n\neq 0}\frac{-T^{|n|}e}{\sqrt{(x-(-1)^{n}(
x_{e}+na))^2+(y-y_{e})^2+(z-z_e)^2}}
\\\nonumber  -\frac{T^{|n|}e}{\sqrt{(x-x_{e})^2+(y-(-1)^{n}(
y_{e}+nb))^2+(z-z_e)^2}}\nonumber
\end{eqnarray}

$V_e^{(im)}$ scales like (1/$r$). It is therefore a general
solution of Laplace equation, however different from the
particular one continuous at the interface between the wire and
its outside medium, $V^{(\epsilon_w \neq \epsilon_m)}_e$.
Nevertheless, as shown below, the discontinuity of $V_e^{(im)}$ at
the boundaries is fairly small, $V^{(\epsilon_w \neq
\epsilon_m)}_e$ being obtained by continuous prolongation of
$V_e^{(im)}$.

With the image representation, the evaluation of the electron-hole
interaction potential requires the inclusion of three
contributions: the direct electron-hole Coulomb interaction, the
interaction between the hole and the electron's images, and the
one between the electron and the hole's images. The electron-hole
interaction potential then reads
\begin{eqnarray}
V^{(im)}(z_e-z_h)= \left< f_{h,1}\left| \bigotimes \left< f_{e,1}\left|
\frac{-e^2}{\sqrt{(x_{e}-x_{h})^2+(y_{e}-y_{h})^2+(z_e-z_h)^2}}\right|
f_{e,1} \right>\bigotimes \right| f_{h,1}\right>\label{eq6}\\
+e\left< f_{h,1}\left| \bigotimes \left< f_{e,1}\left|V^{(im)}_{
e_{in}}(x_{h},y_{h},z_h)-V^{(im)}_{ h_{in}}(x_{e},y_{e},z_e)\right| f_{e,1}
\right>\bigotimes \right| f_{h,1}\right>\nonumber
\end{eqnarray}
which is much simpler to evaluate than (\ref{V_polaire}) since it reduces to
a fourth integration of a rational fraction over a finite domain. Moreover,
as in section I, $V^{(im)}$ is very well approximated by
\begin{equation}\label{eq7}
V_{eff}^{(im)}(z_e-z_h)=-A_{im}\frac{e^{2}}{|z_e-z_h| +b_{im}^{*}}
\end{equation}

The image representation is hereafter applied to evaluate
polydiacetylenes $X_0$ binding energy and Bohr radius. As
previously mentioned, PDA's are excellent quantum wires thanks to
the very regular confinement potential of the monomer crystal
surrounding the chains. In the particular case of poly-3BCMU red
chains, single chain spectroscopy has revealed that red chains
optical excitation, the so-called red exciton, has a center of
mass motion described by a one dimensional band exhibiting the
1/$\sqrt{E}$ energy singularity at the band edge \cite{Dubin2002}.
This undoubtly indicates that this excitation is an excellent
quasi-1D Wannier exciton, which is furthermore highly bound
($E_b$=0.6 eV), with an inferred small Bohr radius ($a_{1D}$=15
$\AA$) \cite{Horvath, Moller}, as expected for quasi-1D $X_0$.

\section{Application to poly-3BCMU chains}

The geometrical cross section of the poly-3BCMU chain is about 4x3
$\AA^2$, using the Van der Waals radius of Carbon atoms
\cite{Enkelmann,Schott_PC}. However, electronic wavefunctions are
not strictly confined within this volume and slightly larger
dimensions are probably appropriate, so the calculations are also
carried out with a 7x5 $\AA^2$ chain cross section. The chain
dielectric constant $\epsilon_w$ has not been measured, but should
be around 10 $\epsilon_0$ like polyacetylene \cite{epschaine}
while the monomer dielectric constant $\epsilon_m$=2.5
$\epsilon_0$ \cite{epsilonm}. Poly-3BCMU chains $X_0$ effective
mass being $\approx$ 0.1 $m_{0}$ ($m_{0}$ denotes the free
electron mass) \cite{Lecuiller2002}, the three dimensional binding
energy and Bohr radius of this excitation are equal to 25 meV and
36 $\AA$ respectively.

\begin{figure}
\includegraphics[width=8cm]{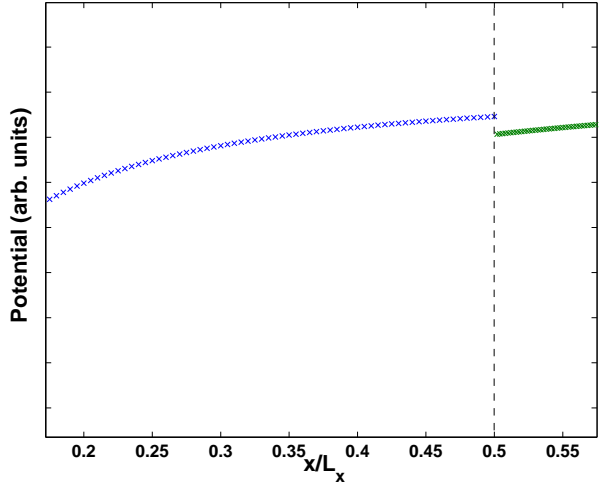}
\caption{Potential $V_{e}^{(im)}$ created along (Ox) by an electron confined
at (0,0,0) within a poly-3BCMU chain. $L_x$=4 $\AA$ is the chain extension
along (0x). The dashed line materializes the boundary between the chain and
its outside medium where the potential discontinuity is 5 $\%$.} \label{fig1}
\end{figure}

Let us quantify the error on the electron-hole interaction
potential for the $X_0$ when calculated by the image
representation. To do so we first evaluate the discontinuity of
$V_{e}^{(im)}$ at the boundary between the poly-3BCMU chain and
the monomer crystal. In Figure \ref{fig1} the potential created
along (Ox) by an electron confined in the center of the chain is
presented. As expected, $V_e^{(im)}$ does not fulfill the boundary
conditions, however its discontinuity at the interface between the
chain and the monomer is only 5 $\%$. In general, $V_e^{(im)}$
discontinuity at the boundary never exceeds 10 $\%$ for any
position of the confined electron. Consequently $V_e^{(im)}$ is a
good approximation of $V_{e}^{(\epsilon_c \neq \epsilon_m)}$.
Moreover, to calculate $V^{(im)}$ (equation (12)), the
electron-hole interaction potential is integrated on the section
of the chain. As shown above, $V_{e}^{(im)}$ only differs from
$V_{e}^{(\epsilon_c \neq \epsilon_m)}$ at the boundary
neighborhood, elsewhere within the chain
$V_{e}^{(im)}$=$V_{e}^{(\epsilon_c \neq \epsilon_m)}$. The
contribution of the boundary (0.4$<|x/L_x|<$0.5 and
0.4$<|y/L_y|<$0.5) in the integration being 5 $\%$, the
calculation of the electron-hole interaction potential by the
image method is then only a $\approx$ 1 $\%$ approximation. This
procedure is here preferred to study poly-3BCMU red chains $X_0$
properties within a well controlled computation.

\begin{figure}
\includegraphics[width=8cm]{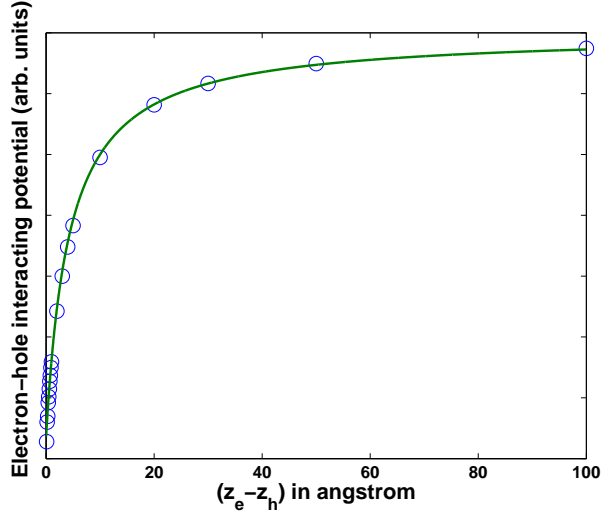}
\caption{Open circles: electron-hole interaction potential for the
$X_0$ given by the image method (equation (\ref{eq6})) for R=0.6,
$L_x$=4 $\AA$ and $L_y$=3 $\AA$. Line: Associated effective
potential (equation (\ref{eq7})) for $A_{im}$=14 and $b^{*}_{im}=
4.3 \AA$.}\label{fig2}
\end{figure}

For a poly-3BCMU chain embedded in its single crystal monomer
marix, $V^{(im)}$ and its associated effective potential
$V_{eff}^{(im)}$ are presented in Figure \ref{fig2}. For
$A_{im}$=14 and $b^{*}_{im}$= 4.3 $\AA$ the effective image
potential is shown to reproduce accurately $V^{(im)}$. For
$b^{*}_{im}$= 4.3 $\AA$, the resolution of equation (\ref{eq4})
yields ($\lambda_{im}$,$\beta^{*}_{im}$)=(0.6,0.87). The binding
energy and Bohr radius of the poly-3BCMU $X_0$ are then deduced
equal to 900 $\pm$ 9 meV and 22 $\pm$ 0.2 $\AA$ respectively,
i.e., in close agreement with measured values. The $X_0$
properties are then well reproduced starting from the most
probable chain conformation. This geometrical factor governs the
calculated $E_b$ and $a_{1D}$, the stronger the confinement is,
the larger $E_b$ and $a_{1D}$ are. Indeed, starting with a
poly-3BCMU chain with a rectangular section of 7x5 $\AA^2$, the
$X_0$ Bohr radius and binding energy are $a_{1D}$=17 $\pm$ 2 $\AA$
and $E_b$=700 $\pm$ 8 meV, i.e., closer to the measured values.
The lateral confinement of PDA chains could thus be slightly wider
than the extension of their Carbon atoms, of the order of five
Angstroem. For such small lateral confinement the validity of our
treatment is questionable; the chain could also be modelled as a
linear chain of atoms in order to investigate its optical
excitation. Nevertheless, from an experimental point of vue,
poly-3BCMU red chains $X_0$ has been evidenced as a quasi-1D
Wannier exciton \cite{Dubin2002}, the corresponding formalism is
then here developed. In regard to our analysis, it seems very
likely that the strong confinement of PDA chains plays a crucial
role in the understanding of their optical properties, the strong
electronic correlations being the other fundamental parameter
which influence should be quantitatively studied.

\section{Conclusion}

The influence of the polarizability of a quantum wire outside
medium on quasi-1D excitonic parameters has been investigated. The
exact electron-hole Coulomb interaction potential for the ground
state exciton is first presented. It is governed by the lateral
confinement and yields the inferred exciton binding energy and
Bohr radius. The exact interaction potential numerical evaluation
being complex, we propose an approximation based on the confined
carriers image charges representation. When applied to the
specific case of PDA chains, our procedure demonstrates its
accuracy. The calculation of the electron-hole interaction
potential for poly-3BCMU red chains is shown to exhibit a 1 $\%$
precision. Furthermore, starting with the chain's most probable
conformation, i.e., with a section of $\approx$ 7x5 $\AA^2$, the
deduced ground state exciton binding energy and Bohr radius are in
good agreement with the measured values. The only relevant
parameter of our model being the cross-section of the chain, we
conclude that the large binding energy and the small Bohr radius
of PDA excitons can both be understood in terms of the strong
confinement of PDA chains, i.e., deriving from the excellent
quasi-one dimensional character of these organic semiconductors.

The authors are very thankful to Monique Combescot for helpful discussions and corrections of the present manuscript.\\

\noindent

$^{1}$: francois.dubin@uibk.ac.at (present adress: Institut für
Experimentalphysik, Universität Innsbruck, Technikerstrasse 25,
A-6020 Innsbruck)

\end{document}